\documentclass[aps,prl,showkeys,floatfix,twocolumn,bibnotes,nofootinbib,superscriptaddress]{revtex4-2}
\usepackage{graphicx,amsmath,amssymb,hyperref,natbib,slashed}
\usepackage[dvipsnames]{xcolor}
\usepackage{xspace}

\usepackage[utf8]{inputenc}

\definecolor{burgundy}{rgb}{0.5, 0.0, 0.13}

\newcommand{\app}{Appendix}

\newcommand{\eVdist}{\kern-0.06em}

\newcommand{\pk}{\mathcal{P}(k)}
\newcommand{\pkr}{\mathcal{P}_\mathcal{R}(k)}

\newcommand{\be}{\begin{equation}}
\newcommand{\ee}{\end{equation}}
\newcommand{\bea}{\begin{eqnarray}}
\newcommand{\eea}{\end{eqnarray}}

\begin{document}
\hfill IPPP/25/40   

\title{Updated constraints on the primordial power spectrum at sub-Mpc scales}

\newcommand{\AddrOslo}{%
Department of Physics, University of Oslo, Box 1048, N-0316 Oslo, Norway}
\newcommand{\AddrDurham}{%
Institute for Particle Physics Phenomenology, Department of Physics, Durham University, Durham DH1 3LE, U.K.}
\vspace*{0.1cm}

 \author{Torsten Bringmann}
 \email{torsten.bringmann@fys.uio.no}
\affiliation{\AddrOslo}

  \author{Djuna Croon}
 \email{djuna.l.croon@durham.ac.uk}
 \affiliation{\AddrDurham}

  \author{Sergio Sevillano Mu\~{n}oz}
 \email{sergio.sevillano-munoz@durham.ac.uk}
 \affiliation{\AddrDurham}

\begin{abstract}
The primordial power spectrum of matter density perturbations contains highly valuable information about new 
fundamental physics, in particular cosmological inflation, but is only very weakly constrained observationally for small 
cosmological scales
$k\gtrsim 3\,{\rm Mpc}^{-1}$. We derive novel constraints, $\mathcal{P}_\mathcal{R}(k)\lesssim 5\cdot10^{-6}$ over
a large range of such scales, from the formation of ultracompact minihalos in the early universe. 
Unlike most existing constraints of this type, our results do not rest on the assumption that dark matter 
can annihilate into ordinary matter.
\end{abstract}


\maketitle

\section{Introduction}%
The power spectrum of primordial density fluctuations, $\pk$, is a key quantity in cosmology. It describes the initial 
conditions of the Universe right after the big bang, tiny perturbations in an almost homogeneous plasma
that eventually evolved into all large-scale cosmological structures that we can see today.
Exquisite observations of the cosmic microwave background radiation (CMB)~\cite{Planck:2018vyg} and 
large-scale structure, in particular through the Lyman-$\alpha$ 
forest~\cite{Bird:2010mp,Fernandez:2023grg}, tightly constrain $\pk$ on length scales larger than $\sim1$\,Mpc. 
These observations are consistent with the simplest models of cosmological inflation~\cite{Martin:2013tda,Achucarro:2022qrl}, 
which represents a remarkable success for our leading explanation of how those primordial perturbations were 
seeded initially. 
$\pk$ is thus not only crucial for understanding the universe at large scales. It also contains invaluable
information about fundamental new physics, connected to much higher energy scales than what is accessible by
earth-bound experiments.

Fully exploiting this potential would require to extend our knowledge of $\pk$ to sub-Mpc scales, but the heap of 
information that 
may be buried there remains largely inaccessible at present. At the smallest scales, in particular,
the non-observation of primordial black holes (PBHs) provides our only probe of $\pk$; this leads to rather
mild constraints, about seven orders of magnitude weaker than what is observed at large 
scales~\cite{Josan:2009qn, Carr:2020gox, Bird:2022wvk}. 
Ultracompact minihalos (UCMHs) form much more frequently from rare density fluctuations 
in the early universe than PBHs~\cite{Ricotti:2009bs,Berezinsky:2003vn,Bringmann:2011ut}.
Assuming that dark matter (DM) can pair-annihilate into standard model particles, like in the case of 
the much-discussed weakly interacting massive particle (WIMP) 
explanation of  DM~\cite{Cirelli:2024ssz,Balazs:2024uyj}, 
this translates into very stringent small-scale constraints on $\pk$~\cite{Scott:2009tu, Josan:2010vn, Bringmann:2011ut,Yang:2013dsa,Nakama:2017qac,Blanco:2019eij,Zhang:2021mth,FrancoAbellan:2023sby, Gouttenoire:2025wxc}.
Future pulsar timing array (PTA) observations of UCMH transients might constrain $\pk$ independent of this 
assumption~\cite{Clark:2015sha,Dror:2019twh}. 
Later forming DM halos may also lead to various constraints at sub-Mpc scales, subject to 
assumptions about late-time evolution and properties of these objects~\cite{%
Yoshiura:2019zxq, Furugori:2020jqn, 
Yoshiura:2020soa, 
Abe:2021mcv, 
Abe:2022swi, 
Ando:2022tpj, 
deKruijf:2024voc, 
Ando:2024ghr%
}.
Finally, spectral distortions of the CMB~\cite{Chluba:2012we} and the induced gravitational wave 
background~\cite{Inomata:2018epa,Byrnes:2018txb} provide complementary constraints directly on the linear, 
or only mildly non-linear, power spectrum at scales down to roughly one pc.

UCMHs are only one example of extended DM objects (EDOs) that may have existed since primordial times.
Recently, in fact, the possibility of  more general macroscopic DM configurations, very compact yet much larger in 
size than PBHs, has attracted 
attention~\cite{Croon:2020wpr,Bai:2020jfm,Croon:2020ouk,Croon:2022tmr,Croon:2024rmw}.  
Even if only interacting gravitationally, EDOs would typically constitute gravitational lenses and accrete ordinary, 
baryonic matter; they may also disrupt other gravitational structures such as 
wide binaries. As it turns out, these effects lead to stringent 
constraints on their cosmological abundance, largely independent of specific assumptions about the 
EDO genesis or the microscopic nature of DM (see Ref.~\cite{Croon:2024jhd} for an up-to-date overview).

Here we combine the idea of using UCMHs to constrain the primordial power spectrum with these
recent developments in the more general EDO context. We present new and significantly extended  limits
on EDOs, and apply them to the more specific case of UCMHs.
In combination, this allows us to improve existing constraints on $\pk$ by orders of magnitude, 
over a wide range of sub-Mpc scales. 
Such tightened small-scale constraints will not only be decisive in further narrowing down the microphysics governing
inflation~\cite{Peiris:2008be, Adshead:2008vn, Josan:2010cj, Aslanyan:2015hmi,Hertzberg:2017dkh}, 
but also put highly relevant bounds
on other mechanisms generating excess power on small scales, like dark sector phase 
transitions~\cite{Gouttenoire:2025wxc}. {Last but not least, our constraints have far-reaching consequences
for the possibility of PBHs with $M\gtrsim 0.1\,M_\odot$ constituting even a tiny fraction of DM.}

\section{UCMH formation and properties}%
Let us consider a region of comoving radius 
$R$ in the early universe, with an average density $\rho$ that is larger than the
cosmological background density $\bar\rho$. For sufficiently large overdensities 
$\delta\equiv (\rho-\bar\rho)/\bar\rho>\delta_\chi^{\rm min}\sim10^{-3}$ during radiation domination, this  region
will collapse into virialized DM objects shortly after matter-radiation equality~\cite{Ricotti:2009bs}. 
Since such large values of $\delta$ correspond to very rare fluctuations, the collapse 
typically happens in isolation and by almost radial 
infall, leading to the formation of highly compact objects, dubbed UCMHs. 
Their mass corresponds to the DM mass within 
the horizon $H^{-1}$ at mode entry,
\be
M_{\rm UCMH}\simeq\frac{4\pi}{3}\left[\rho_\chi H^{-3}\right]_{aH=1/R}=\frac12M_{\rm Pl}^2H_0^2\Omega_\chi R^3\,,
\label{eq:Mhorizon}
\ee
where $a$ is the scale factor, $M_{\rm Pl}$ the Planck mass, $\rho_\chi$ the DM density and 
we introduced $\Omega_\chi\equiv \rho_\chi/\rho_c=0.264$~\cite{Planck:2018vyg}, with $\rho_c$ being the critical 
density.
It was originally argued that accretion would enhance $M_{\rm UCMH}$ by another factor of $a_{\rm eq}/a$ after 
equality, but such a growth could not consistently be reproduced in simulations~\cite{Delos:2017thv} and we
hence conservatively neglect it for most of our work (but discuss this further in the \app).

The density profile inside these objects is very steep, $\rho_{\rm UCMH}(r)\propto r^{-\alpha}$,  with
$\alpha=9/4$ expected for self-similar, spherical 
collapse~\cite{Fillmore:1984wk,Bertschinger:1985pd,Vogelsberger:2009bn,Ludlow:2010sy}. 
While this value of $\alpha$ is realistic for fully isolated overdensities, numerical simulations often find
less spiky profiles~\cite{Diemand:2005vz,Angulo:2016qof,Ishiyama:2014uoa}, even for  
rather early collapse~\cite{Delos:2017thv,Gosenca:2017ybi,Delos:2018ueo}.
We will hence assume $\alpha=3/2$, noting that our limits anyway only weakly depend 
on the exact value of $\alpha$ 
(see \app\ for a more detailed discussion). 
Unlike for the case of DM annihilation, our analysis is also insensitive to the inner cutoff 
radius at which $\rho_\chi$ plateaus because of the non-vanishing velocity distribution or DM depletion due to
self-annihilation~\cite{Bringmann:2011ut}. We take the forming UCMH to extend out to an effective radius
where its density has dropped to three times the background density, 
$R(z)\simeq21 
(1+z)^{-1}(M_{\rm UCMH}/M_\odot)^{1/3}\,{\rm pc}$~\cite{Ricotti:2007au}. The radius within which 
$90\%$ of the mass is enclosed is thus given by $R^{(90)}_{\rm UCMH}=(0.9)^{3/2}R(z_c)$, with 
$z_c$ the redshift of collapse. This translates to 
$\rho_{\rm UCMH}(r)=\rho_{90} \big(r/R^{(90)}_{\rm UCMH}\big)^{-3/2}$ with 
$\rho_{90} \simeq1.4\cdot 10^{-5}(1+z_c)^3\,M_\odot\, {\rm pc}^{-3}$. 

For Gaussian fluctuations, the cosmological abundance of UCMHs is exponentially sensitive to the critical overdensity 
$\delta_\chi^{\rm min}$ for collapse to happen. We follow Ref.~\cite{Bringmann:2011ut} to calculate this value,
very conservatively requiring collapse by $z_c=1200$. For the range of scales that will be relevant for
us, $3\,{\rm Mpc}^{-1}\lesssim k\lesssim 10^5\,{\rm Mpc}^{-1}$,  this corresponds to  
$1.03\cdot10^{-2}\gtrsim \delta_\chi^{\rm min}\gtrsim 2.25\cdot 10^{-3}$. 

\section{Updated EDO limits}%
Recently, gravitational constraints on generic EDOs 
have been derived from the non-observation of micro-lensing events~\cite{Croon:2020ouk,Croon:2020wpr}, 
gravitational wave 
production~\cite{Croon:2022tmr}, the evaporation of wide binaries~\cite{Ramirez:2022mys,Qiu:2024muo}, dynamical 
heating of dwarf galaxies~\cite{Graham:2024hah}, and heating of the CMB due to baryonic 
accretion~\cite{Croon:2024rmw}. A repository of current constraints on EDOs with different mass profiles is 
provided in Ref.~\cite{Croon:2024jhd}. Here we extend bounds from accretion and 
introduce a new constraint from microlensing of the MACS J1149 LS1 ``Icarus'' star.

In previous analyses the accretion constraint was taken to be valid up to masses of
$M = 10^5 M_\odot/\sqrt{f_{\rm DM}}$, where $f_{\rm DM}$ is the DM fraction in EDOs. Beyond 
this the average EDO separation is much smaller than their Bondi radius $R_{\rm B}(z) = GM/c_s(z)^2$,
with $c_s$ being the sound speed, raising concerns about the underlying assumption of isolated objects. A priori, 
exceeding this limit may introduce backreaction effects (potentially underestimating constraints) or lead to overcounting 
the total emitted radiation (potentially overestimating them). However, this latter effect is significantly suppressed, as only 
the ionised matter (heated above $T_{\rm ion} = 1.3\,\rm eV$~\cite{1991ApJ...383..250N}) 
contributes to luminosity. This only occurs 
within the ionisation radius $R_{\rm ion} = R_{\rm B} (T_{\rm CMB}(z)/T_{\rm ion})\,\Theta(z)^{-2/3}$, where $\Theta(z)$ 
accounts for interactions with the background~\cite{Bai:2020jfm,Croon:2024rmw}. 
Since $R_{\rm ion} \ll R_{\rm B}$ stays well 
below the average EDO separation across relevant redshifts, 
the isolated object approximation remains 
a conservative and valid basis for extending constraints to higher-mass 
EDOs.

As discussed in more detail in the \app,
observations of highly magnified background stars behind galaxy clusters yield a novel independent bound on 
EDOs. 
In favourable lensing configurations, where a distant star is strongly amplified by a cluster caustic, the magnification 
hinges 
on the smoothness of the cluster's mass profile. If DM consists of compact objects, it fragments the caustic into a 
network of microcaustics, modifying the lensing signal. Lensed observations of MACS J1149 LS1 
can therefore be used to constrain the DM fraction in compact objects~\cite{Oguri:2017ock,Diego:2017drh,Croon:2025yfj}.
In an extended macrolens, the effective Einstein radius of a point lens at distance $D_{L}$ and Einstein angle $\theta_{E}$ is
$\bar R_E = \sqrt{\mu_t}\,\theta_E\,D_L,$ 
which for MACS J1149 LS1 is boosted by at least $\mu_t\sim15.4$ \cite{Oguri:2017ock}. Lenses behave as point-like 
for a physical size~\cite{Croon:2020wpr,Croon:2020ouk}
\begin{equation}
  R_{\rm 90}\lesssim\bar R_E
\approx2.0\times10^6\bigl(M/M_\odot\bigr)^{1/2}R_\odot\,,
\end{equation}
using $\theta_E = 1.8 \times 10^{-6} (M/M_\odot)^{1/2}\rm arcsec $ and $ D_L = 6.4 \, \rm kpc \, arcsec^{-1}$ for this 
system \cite{Oguri:2017ock}.
Following~\cite{Croon:2025yfj}, we rescale the point-like constraints by an extended lens
efficiency $\varepsilon$. The finite source size enters the calculation as in Refs.~\cite{Oguri:2017ock,Croon:2025yfj}.

\begin{figure}[t]
\includegraphics[width=\columnwidth]{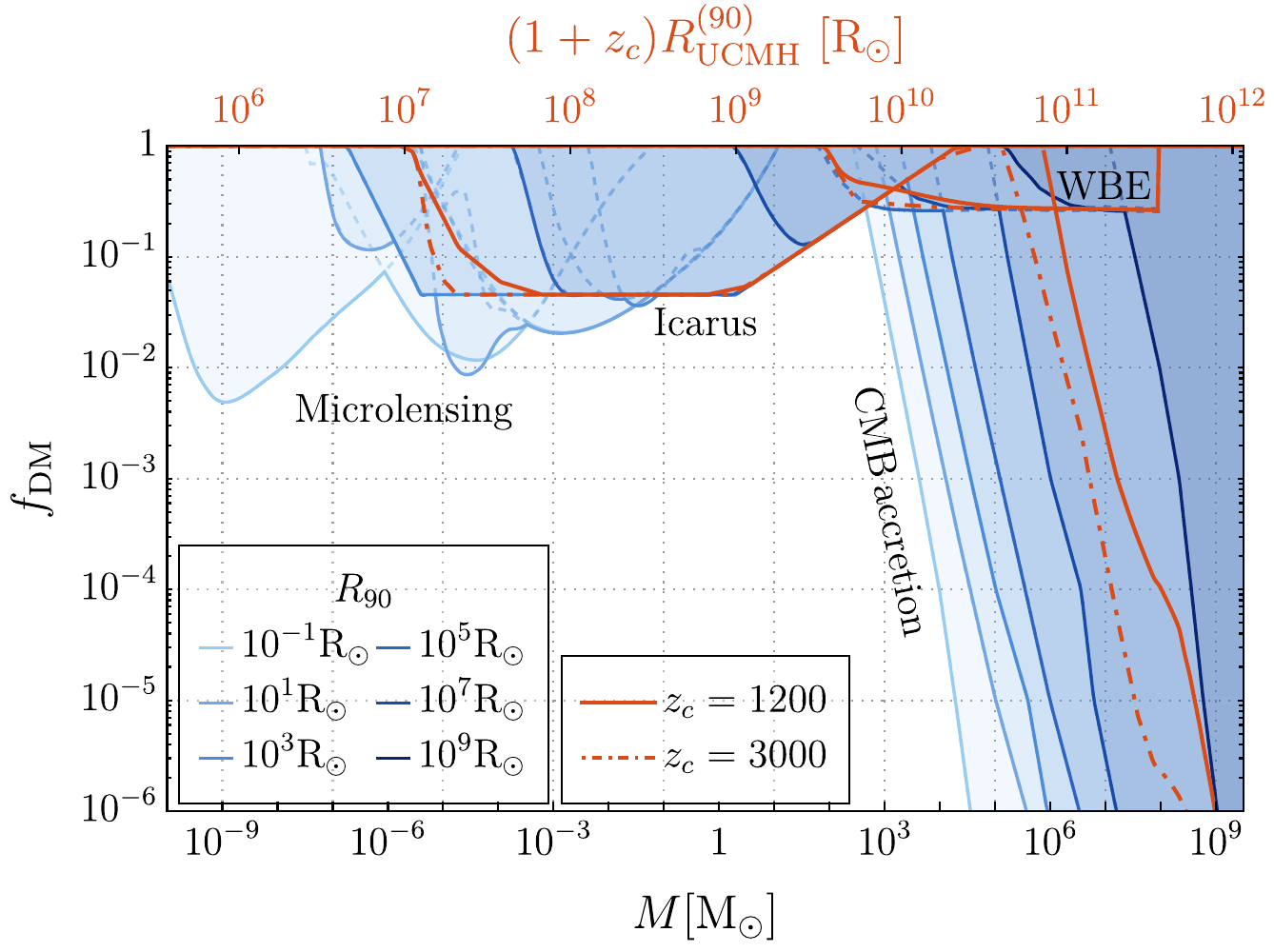}
\caption{%
Present bounds on the DM fraction $f_{\rm DM}$ comprised of EDOs  (blue), with fixed radii as stated in the 
legend, and UCMHs (red). 
Constraints deriving from lensing observations of the Icarus star (MACS J1149 LS1) and from
CMB heating due to baryonic accretion are described in detail in the text.
We further apply Galactic microlensing~\cite{Croon:2020wpr,Croon:2020ouk} and wide binary 
evaporation constraints~\cite{Ramirez:2022mys}.
}
\label{fig:abundance_limits}
\end{figure}

In Fig.~\ref{fig:abundance_limits} we plot the constraints on the EDO abundance  discussed above
(blue contours), for various fixed values of the size $R_{90}$ of these objects. We choose a density profile $\rho_{\rm 
EDO}\propto r^{-\alpha}$ with $\alpha=3/2$, stressing however that 
these constraints are not very sensitive to the precise value of $\alpha$ 
(see also the \app ) 
and mostly depend on the underlying relation between mass and 
$R_{90}$~\cite{Croon:2024rmw,Croon:2024jhd}.
In red, we overlay the constraints on objects which satisfy the UCMH mass-radius relation 
$R^{(90)}_{\rm UCMH}\propto M^{1/3}$. 
We note that UCMHs are not sufficiently compact to be probed by Galactic microlensing, 
see also the discussion below,
and that CMB accretion constraints by far dominate at high masses.

\begin{figure*}[t]
\includegraphics[width=0.95\textwidth]{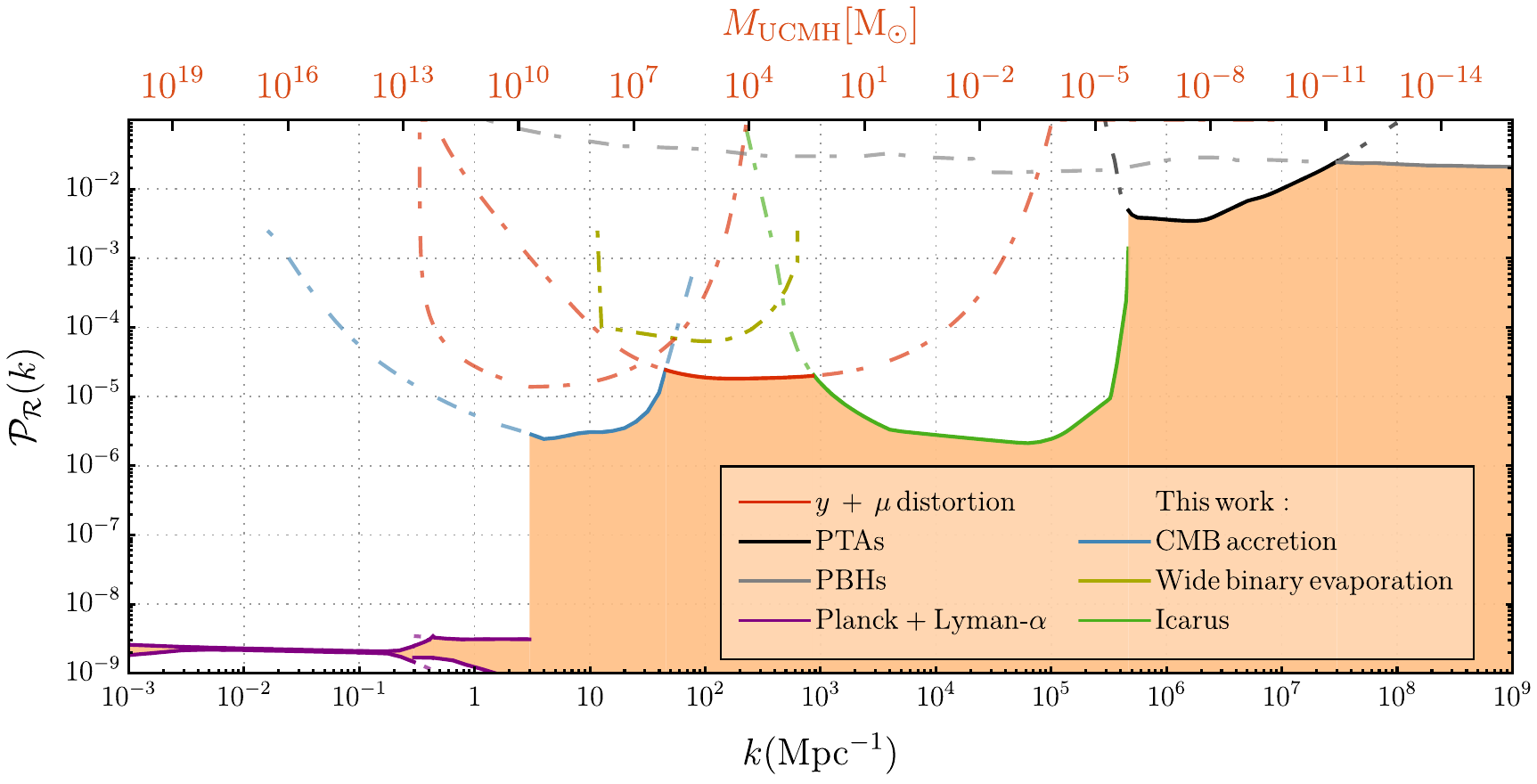}
\caption{%
Shaded regions show the presently allowed strength of a locally scale-invariant primordial curvature power spectrum 
$\pkr$, as a function of co-moving scale $k$. Our newly derived limits from UCMHs due to CMB accretion (blue),
wide binary evaporation (olive) and  ICARUS lensing observations (light green) are discussed in detail in the text. 
We also present updated constraints deriving from PTAs (black), as well as existing CMB $y$ and $\mu$ distortion
constraints applied to the form of $\pkr$ assumed here (red). For comparison, we further show existing 
constraints at large scales (purple) deriving from CMB and Lyman-$\alpha$ 
observations~\cite{Planck:2018vyg,Chluba:2015bqa,Green:2020jor}, as well as 
PBH bounds~\cite{Carr:2020gox} that extend to very small scales (grey). 
}
\label{fig:constraints_summary}
\end{figure*}

\section{UCMH constraints on the power spectrum}%
In the previous section we presented updated constraints on the EDO abundance. By applying those to UCMHs,
we now show how this translates to constraints on the curvature 
power spectrum $\pkr$. Assuming Gaussian fluctuations,
the probability for a region of size $R$ to eventually collapse into a UCMH is given by
\be
\label{eq: beta(r)}
    \beta=\frac{1}{\sqrt{2 \pi} \sigma_{\chi,H}}\int^{\delta^{\rm max}_{\chi}}_{\delta^{\rm min}_{\chi}}\exp\left[-\frac{\delta_\chi^2}{2 \sigma^2_{\chi,H}}\right] {\rm d}\delta_\chi\,,
\ee
which depends only very weakly on the upper integration limit 
$\delta^{\rm max}_{\chi}\simeq\mathcal{O}(1)\gg \delta^{\rm min}_{\chi}$ corresponding to the formation of a PBH 
rather than an UCMH. Here, the mass variance $\sigma_{\chi,H}$ at horizon crossing of the scale $R$  is computed by 
convolving the power spectrum with a top-hat window function~\cite{Bringmann:2001yp,Bringmann:2011ut}. For scales 
$k_R=1/R$ entering during radiation domination, it takes the form
\be
\label{eq:sigh}
 \sigma_{\chi,H}^2 = \! \!\int_0^\infty \!\!\!{\rm d}x\,\frac{\left(\sin x - x \cos x\right)^2}{x^3}\mathcal{P}_\mathcal{R}(x/R)\,T_\chi(x/\sqrt{3})\,,
\ee
with the transfer function $T_\chi$ given explicitly in Eq.~(A10) of Ref.~\cite{Bringmann:2011ut}.
For most functional forms of $\pkr$, this integral
is strongly dominated by $x\sim1$, corresponding to scales $k\sim k_R$. 
{For ease of comparison with the literature, 
we will present our constraints in terms of a power spectrum that is {\it locally scale-invariant}, 
$\mathcal{P}_\mathcal{R}(k) = \mathcal{P}_\mathcal{R}(k_\mathrm{R})$ for $k\sim k_R$, leading to
$ \sigma^2_{\chi,{\rm{H}}}(R)\simeq 0.91\,\mathcal{P}_\mathcal{R}(k_\mathrm{R})$.}
We stress,
however, that Eq.~(\ref{eq:sigh}) can, without further complications, also 
be used to constrain power spectra of arbitrary shapes.
Non-Gaussianity
would modify the tail that sets $\beta$, and hence the inferred
$\mathcal{P_R}(k)$; the effect is model-dependent but generically
milder than for PBHs, since UCMH formation samples a far less
extreme part of the distribution ($\delta^{\min}_\chi\sim10^{-3}$).

Neglecting the accretion of baryonic matter,  the final UCMH abundance is given as 
$\Omega_{\rm UCMH}=\Omega_\chi\beta$. 
Requiring this to be smaller than the limits on $f_{\rm DM}$ reported above leads to the constraints on $\pkr$ shown in 
Fig.~\ref{fig:constraints_summary}. 
Note that for our default choice of $z_c=1200$, the Icarus constraint does not apply for $k\gtrsim3\cdot 10^4\,{\rm Mpc}^{-1}$, 
cf.~Fig.~\ref{fig:abundance_limits}, because $R^{(90)}_{\rm UCMH}$ then becomes smaller than the Einstein radius
of the UCMH. For these scales, we instead adopt the smallest $z_c>1200$ that still results in a 
constraint at a given $k$ (see also Fig.~3 in~the Appendix). 

\section{Further constraints}%
For this range of scales, it has previously been pointed out that
also the lack of Compton-$y$ and $\mu$ distortions of the CMB black-body spectrum constrains scalar perturbations.
We follow Ref.~\cite{Chluba:2015bqa} in computing $y$ and $\mu$, 
assuming again $\pkr\simeq const.$ around the pivot scale $k_R$. For concreteness, 
we assume that  $\pkr$ is nonzero only for the range $1/3\lesssim k/k_R\lesssim 3$.
Comparing the result to the $2\sigma$ limits of $y<1.5\times10^{-5}$ and $\mu<9\times10^{-5}$ from
FIRAS observations~\cite{Fixsen:1996nj} then leads to the constraints shown as red lines in 
Fig.~\ref{fig:constraints_summary}.
We caution, however, that $\mu$ and $y$ in general receive contributions from a significantly larger range of scales than 
$\sigma_{\chi,H}$ in Eq.~(\ref{eq:sigh}), which necessarily makes any  comparison with UCMH 
limits somewhat model-dependent 
(see \app\ for a more detailed discussion). 

Another relevant constraint that is independent of UCMH formation derives from the fact that scalar 
perturbations produce a stochastic gravitational wave background at second order in perturbation
theory~\cite{Ananda:2006af,Baumann:2007zm}. 
We compute the resulting power spectrum in gravitational waves, $\mathcal{P}_h$,
semi-analytically by adopting an averaging procedure as advocated 
in Refs.~\cite{Kohri:2018awv,Inomata:2016rbd}, 
noting that the innermost integral is dominated by contributions from modes well inside the horizon.
We assume $\pkr$ to be scale-invariant over the same range of 
scales as for CMB distortions.
We then compute the resulting fractional energy density of gravitational waves today, 
$\Omega_{\rm GW}h^2(k)=(\Omega_r h^2)\frac{1}{24} ({k}/a H)^2 \mathcal{P}_h(k,t)$,  
where the r.h.s.~is evaluated at any time $t_k\lesssim t \lesssim t_{\rm eq}$ 
and the factor of $\Omega_r h^2=4.15\cdot 10^{-5}$~\cite{Fixsen:1996nj} 
accounts for the red-shifting of $\Omega_{\rm GW}$ during matter domination.
We display the resulting limits on $\pkr$ as black lines in 
Fig.~\ref{fig:constraints_summary}, resulting from the requirement of $\Omega_{\rm GW}h^2$
not exceeding the NANOGrav data~\cite{NANOGrav:2023hvm} at 99\% C.L in either of the frequency bins.

We also include a compilation of PBH bounds~\cite{Carr:2020gox} in the figure, 
which remain the strongest constraints for scales $k\gtrsim10^6\,{\rm Mpc}^{-1}$.
At large scales, finally, the power spectrum is  
well measured by observations of the CMB temperature fluctuations
and the Lyman-$\alpha$ forest. Here we show results
from Ref.~\cite{Green:2020jor} for the former, based on Planck data~\cite{Planck:2018jri}, and from 
Ref.~\cite{Fernandez:2023grg} for the latter.

\section{Discussion}%
We have demonstrated that the formation of UCMHs leads to highly competitive constraints on $\pkr$ 
down to scales $k\sim10^5$\,kpc. Remarkably, these constraints do not depend on interactions between DM
and ordinary matter and hence apply to any collision-less and sufficiently cold DM candidate.
The underlying assumption is thus the same as for the complementary constraints shown in 
Fig.~\ref{fig:constraints_summary}, namely that no {\it additional} physical processes wash out density 
perturbations on sub-horizon scales.
For DM particles that have been in thermal equilibrium in the early universe, e.g., these constraints apply below the scale 
of free streaming or kinetic decoupling, $k<k_D$, 
with $k_{D}\sim 10^{5}-10^8\,{\rm Mpc}^{-1}$ for canonical WIMPs~\cite{Bringmann:2009vf}.
We emphasize that we have
followed a very conservative approach by restricting our analysis to objects that have fully collapsed very
early, before a redshift of $z_c=1200$. Notably, this considerably simplifies the analysis as it largely avoids 
complications due to potential changes in the inner density profile
or tidal disruption affecting the survival probability at later formation times. 

Still, even significantly smaller initial density contrasts than considered here would imply perturbations that enter
the non-linear regime well before the onset of standard structure formation.  While UCMHs forming at 
$z_c\lesssim1000$ no longer efficiently contribute to the CMB accretion bounds, both wide binary evaporation
and the Icarus observation therefore generally lead to significantly stronger limits than what we reported in 
Fig.~\ref{fig:constraints_summary}. Due to the aforementioned effects, however, such more aggressive 
limits should be considered less robust. To a much lesser degree, our constraints are affected by other 
assumptions about the UCMH properties. Taking into account a potential mass growth due to baryonic accretion,
in particular, would lead to somewhat stronger bounds, while assuming a shallower profile (as expected for 
later collapse) would result in slightly weaker constraints. 
For a more detailed discussion, 
we refer to the \app. 

Before concluding, let us briefly discuss further, complementary probes of the small-scale power spectrum. 
To start with, microlensing is effective when the physical size of the lens is smaller than the Einstein radius 
$ R^{(90)}_{\rm UCMH} \leq R_{E,\text{max}} = \sqrt{{G M D_S}/{c^2}} $,  
where $ D_S $ is the distance to the source and $M$ the lens 
mass. 
For Galactic microlensing, this formally implies $M>10^{{17}}\, \rm M_{\odot}$, 
i.e.~far above the entire Galaxy's mass. 
This explains why there are no UCMH microlensing constraints in Figs.~\ref{fig:abundance_limits} 
and \ref{fig:constraints_summary} but,
if a greater UCMH compactness is assumed, future microlensing 
observations can lead to constraints on $\mathcal{P}(k)$~\cite{Delos:2023fpm}. 
In contrast, the larger source distance and effective boost in $R_{E}$ make highly magnified stars such as Icarus an 
effective probe.
Future observations of gravitational wave lensing with both sources and lenses at cosmological distances 
may also lead to constraints on UCMH \cite{Oguri:2020ldf,Kim:2025njb}.
Astrometric microlensing, finally, can be detectable even if the magnification is small and may provide
a further future probe \cite{Li:2012qha}. 

Recently, Ref.~\cite{Qin:2025ymc} translated the Planck optical depth into limits on $\mathcal P_{\mathcal R}(k)$ at low 
$k\sim3\!-\!50\ \mathrm{Mpc}^{-1}$, using a one-zone reionisation model with fixed ionising efficiency, escape fraction 
and IGM clumping, assuming UCMH survival until $z\sim200$. At higher $k\sim10\!-\!10^{3}\ \mathrm{Mpc}^{-1}$, 
Ref.~\cite{Graham:2024hah} derived dynamical-heating constraints from ultrafaint 
dwarf galaxies complementary to the bounds shown here, however using NFW subhalos instead of the UCMHs 
considered here. Like the microlensing constraints, these rely on subhalo survival without significant tidal disruption or 
baryonic feedback.

Finally, we note that the results in this article may also have implications for isocurvature perturbations, since the 
formation of UCMHs is highly sensitive to non-adiabatic fluctuations on small scales. In particular, improved constraints 
on the small-scale power spectrum may tighten bounds on isocurvature modes that evade CMB limits but still 
source early structure formation.

\section{Conclusions}%
While PBHs currently remain the only probe of the smallest cosmological scales, with $k\gtrsim10^5$\,Mpc$^{-1}$,
there are several independent observations that put much more stringent constraints on the primordial
power spectrum at slightly larger scales. Here we pointed out that very early forming UCMHs can be used to
significantly tighten these bounds, in a way that is both robust and largely independent of the nature 
of DM. 
Notably, this also provides complementary bounds on PBH DM, 
essentially excluding the possibility of even sub-dominant populations, with $f_{\rm PBH}\ll1$, that 
formed from the collapse of initial density perturbations at scales $k\lesssim10^7\,{\rm Mpc}^{-1}$ (corresponding to 
$M_{\rm PBH}\gtrsim 0.1\,M_\odot$). 

Future optical and gravitational wave lensing observations have the potential of even further
improvement. Such probes of the small-scale primordial power spectrum constitute a crucial step towards 
understanding the microphysics governing cosmic inflation, as well as more exotic processes that may 
generate excess power at these scales.


\bigskip
\vfill
 \paragraph*{Acknowledgements.---}%
We thank Kai Schmidt-Hoberg for very useful discussions, leading to improvements of an initial
version of the draft.
TB gratefully acknowledges support through a FRIPRO grant of the Norwegian Research Council (project ID 353561
`DarkTurns').~DC and SSM are supported by the STFC under Grant No.~ST/T001011/1. TB and SSM are grateful to the Mainz Institute for
Theoretical Physics (MITP) of the Cluster of Excellence PRISMA+ (Project ID 390831469),
for its hospitality and partial support during the DMLAND2024 workshop where the idea for this work was born. 
\vfill


\bigskip 
\appendix
\newpage
\section*{Appendix}

Here we complement the discussion in the main text with additional, more  technical information
about the underlying assumptions connecting to UCMH formation and how they affect our results.
We also demonstrate quantitatively how non-UCMH constraints, relating to
$\mu$ distortions of the CMB and gravitational wave production from scalar perturbations, 
are more model-dependent in that they probe a broader range of scales in $\pkr$.
Finally, we provide more details about lensing constraints from highly magnified stars 
like Icarus.

\subsection{UCMH formation}
In order to be as conservative as possible, we so far only considered UCMHs that collapse very early, namely 
by a redshift of 
$z_c=1200$. A collapse at smaller redshifts would require smaller and thus much more likely
initial density contrasts $\delta\gtrsim\delta_c$.
On the other hand, later forming objects are less dense, with $\rho_{90}\propto (1+z_c)^3$, and extend to larger 
radii $R_{\rm UCMH}^{(90)}\propto (1+z_c)^{-1}$. Notably, the latter aspect does not only affect the survival 
probability, but even detection prospects under the (optimistic) assumption that the late-time cosmological 
evolution does not affect abundance and structure of these objects.

\begin{figure}[t]
  \includegraphics[width=\columnwidth]{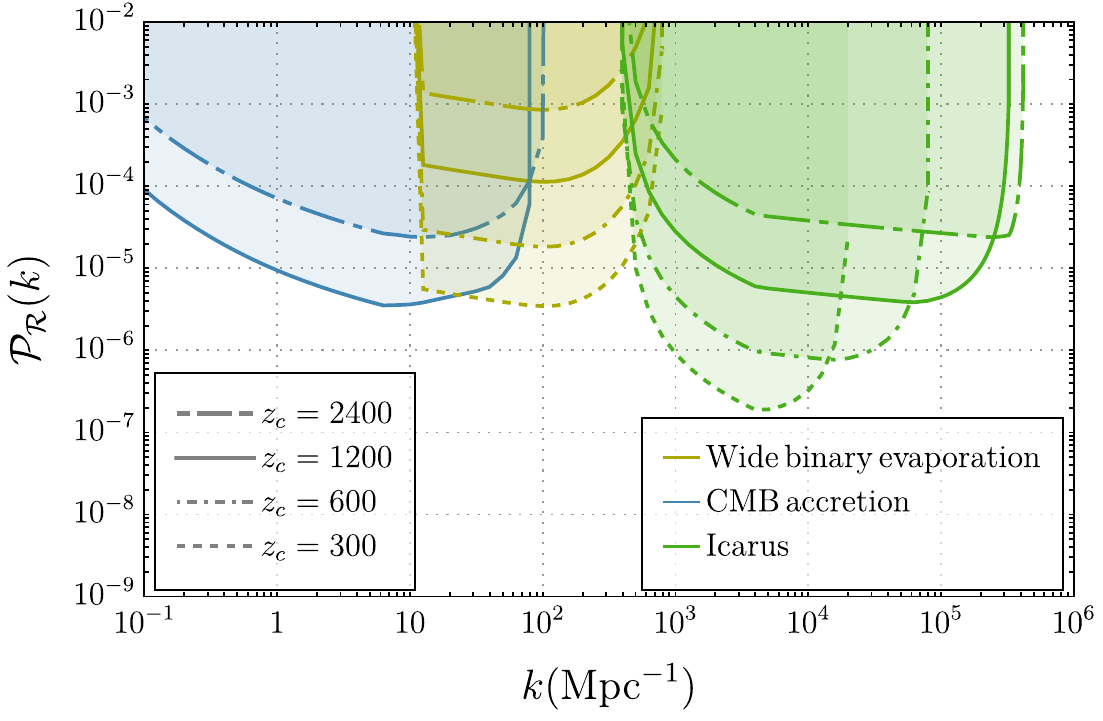}
  \caption{Wide binary evaporation and Icarus constraints on $\pkr$, for various assumed 
  redshifts $z_c$ of collapse. CMB accretion bounds require formation before recombination.
  }\label{fig:redshift_collapse}  \end{figure}

In Fig.~\ref{fig:redshift_collapse} we demonstrate how our limits on $\pkr$ depend on the assumed value of $z_c$, 
noting that CMB bounds
in any case require collapse before recombination. The other constraints, from wide binary evaporation and 
lensing of the Icarus star, clearly show the generally expected behaviour outlined above: while the  
constraint on the power spectrum normalization tightens as a direct consequence of the decrease in $\delta_c$,
the mass range where the constraints apply is reduced as the physical UCMH size increases. 
The actual reduction of the maximally constrained value of $k$ visible in the figure is stronger than 
the naively expected $\propto (1+z_c)$, because the Icarus bound itself is mass-dependent.

In reality, collapse would happen at {\it all} the redshifts indicated in the figure, so the overall constraint on $\pkr$ should in 
principle be taken as the envelope of the curves shown here. However, we re-iterate that these constraints become
less and less robust as $z_c$ is decreased. For late-time collapse, a more realistic computation should thus take into 
account both survival probability and changes in the UCMH density profile due to merger events and other effects
during non-linear structure formation.
This must be based on full numerical simulations and is beyond the scope of this work.
Taking into account {\it earlier} collapse times, as also shown in Fig.~\ref{fig:redshift_collapse},  
is particularly relevant for the Icarus constraint. This is because the microlenses must be 
smaller than their Einstein radius, $1>R^{(90)}_{\rm UCMH}/R_E\propto k^\frac12 (1+z_c)$, 
such that a larger $z_c$ allows to probe larger $k$. At the same time, $R_E\propto \sqrt{M_{\rm UCMH}}$
must remain larger than the stellar radius (see also the discussion further down), corresponding to 
the $z_c$-independent bound $k\lesssim 2\cdot 10^{5}\,{\rm Mpc}^{-1}$.

\subsection{UCMH density profiles and growth by accretion}
\begin{figure}[t]
  \includegraphics[width=\columnwidth]{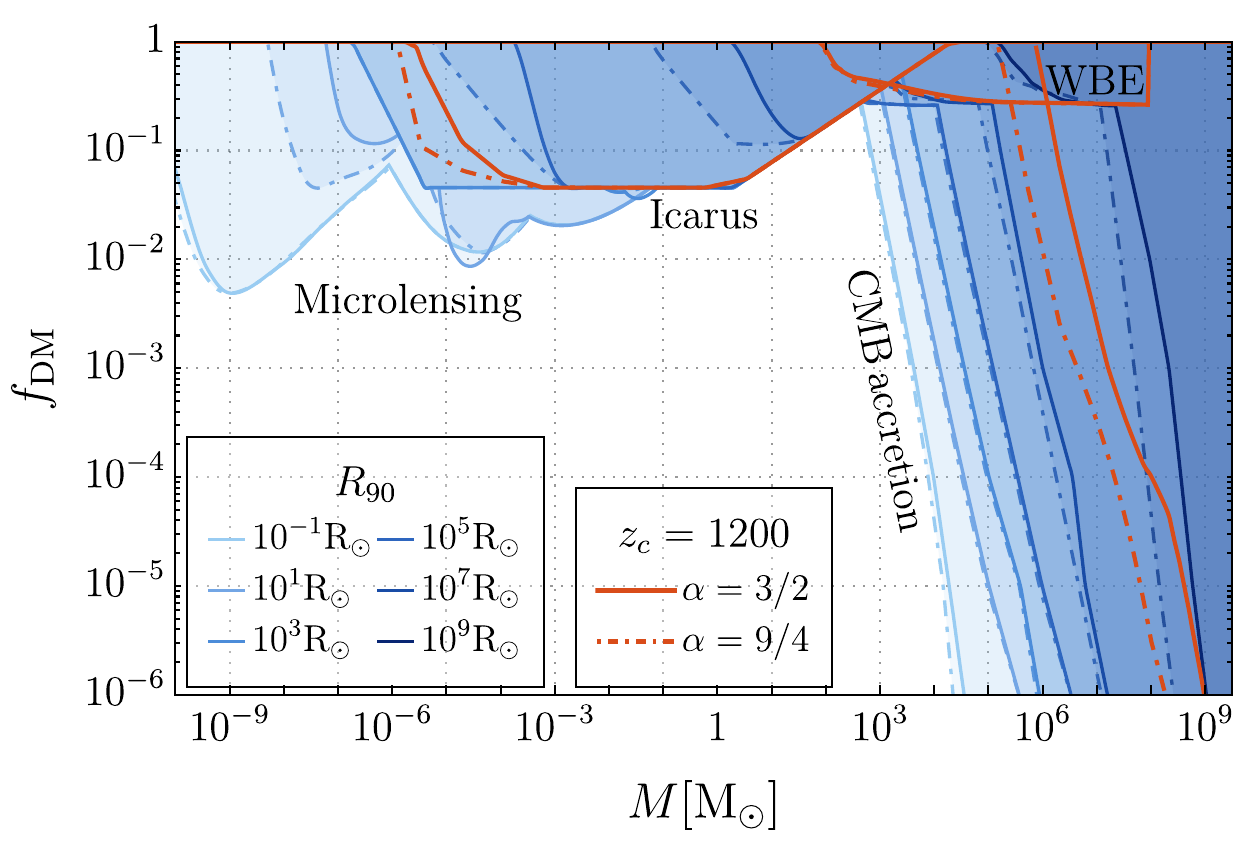}
  \caption{%
  EDO bounds for different density profiles $\rho\propto r^{-\alpha}$, for fixed $R_{90}$ (blue)
  and for UCMHs (red), respectively. The solid lines are the same as in Fig.~\ref{fig:abundance_limits} 
  in the main text. 
  } \label{fig:EDO_bounds_32}
  \end{figure}

As stressed in the main text, UCMHs forming very early and in isolation are expected to obtain very steep
density profiles $\rho\propto r^{-\alpha}$ with $\alpha=9/4$. Once either of these assumptions is not met,
however, somewhat less steep profiles with $\alpha\simeq 3/2$ are more realistic. Also independently of the
UCMH context it is interesting to ask how our general EDO constraints are affected by the density profile 
for $r<r_{90}$. We show this in Fig.~\ref{fig:EDO_bounds_32}, demonstrating that the value of $\alpha$ only has
a relatively mild impact on the limits. 
For both the wide binary evaporation constraint and the CMB accretion constraint,
this remains true as long as the objects under consideration 
remain at least somewhat compact, implying that such limits are dominantly driven by the relation between $M$
and $R_{90}$~\cite{Croon:2024rmw,Croon:2024jhd}. 
The microlensing efficiency for extended objects displays some sensitivity to the density profile, as was 
discussed in Ref.~\cite{Croon:2020wpr}, where the threshold impact parameter 
effectively encodes the efficiency (see in particular their Fig.~3); for the Icarus constraint, see the discussion
in Refs.~\cite{Oguri:2017ock,Croon:2025yfj}.

\begin{figure}[t]
\includegraphics[width=\columnwidth]{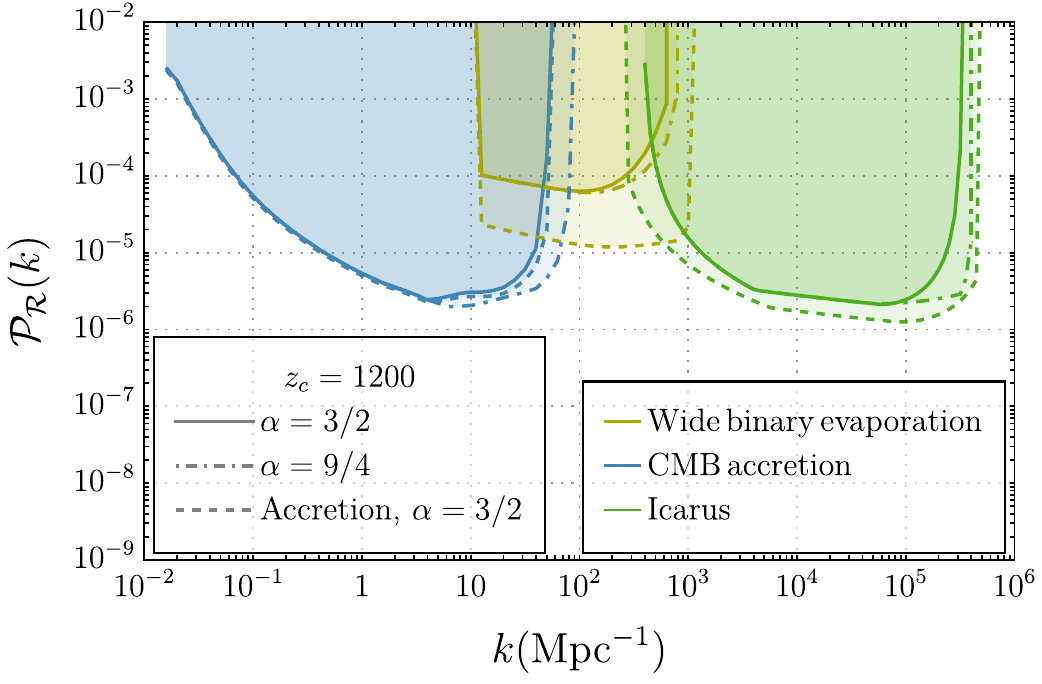}
\caption{%
Comparison of UCMH constraints for different inner density profiles and accretion mechanisms.
Solid lines correspond to the limits presented in Fig.~\ref{fig:constraints_summary} in the main text.
Dash-dotted lines show the same constraints  assuming  a density profile $\rho_{\rm UCMH}\propto r^{-\alpha}$
with $\alpha=9/4$ rather than $\alpha=3/2$. Dashed lines show the effect of a UCMH mass growth due to accretion,
as explained in the text.}
\label{fig:shape_comparison}
\end{figure}

As shown explicitly in Fig.~\ref{fig:shape_comparison},  assumptions about the inner density 
profiles of UCMHs only have a mild impact on our constraints on the primordial power spectrum.
We see that the constraints for steeper density profiles are slightly stronger at small scales, corresponding to the 
greater sensitivity to small masses in Fig.~\ref{fig:EDO_bounds_32}. 

Another aspect entering our limits is how we treat the UCMH mass.
In the main text, we conservatively assume that $M_{\rm UCMH}$  is given by Eq.~(\ref{eq:Mhorizon}), with no 
subsequent growth due to accretion. While this behaviour corresponds to the finding in some simulations, 
other works have argued that $M_{\rm UCMH}$ should grow by a factor of $(1+z_{\rm eq})/(1+z_c)$ during matter 
domination.
We illustrate in Fig.~\ref{fig:shape_comparison} the impact of such a mass growth, 
concluding that the impact on the CMB accretion and Icarus limits on $\pkr$ is noticeable but still relatively small, 
while the impact on the wide binary evaporation limits is significant. In combination, this would imply that 
UCMH constraints outperform CMB spectral distortion constraints over the entire range of scales considered in our 
analysis. The reason why wide binary evaporation is particularly sensitive to the UCMH mass is simply
that the constraint on $\pkr$ scales nonlinearly with the constraint on $f_{\rm DM}$, such that a fractional 
change in a less stringent constraint on $f_{\rm DM}$ affects $\pkr$ more strongly.
We conclude our discussion of the potential impact of a UCMH mass growth by noting that 
a growth $\propto (1+z_c)^{-1}$ can in any case not be the correct prescription for large values of $\pkr$: 
once there is a significant probability $\beta\gtrsim \mathcal{O}(10^{-3})$ for a region of a given size to collapse, 
cf.~Eq.~(\ref{eq: beta(r)}), there is no longer a sufficient amount of unbounded background material in the surrounding 
regions to sustain such a mass growth
(formally resulting in `overclosure constraints' $\Omega_{\rm UCMH}>\Omega_{\rm DM}$ if applying the 
$\propto (1+z_c)^{-1}$  scaling also in this regime).

\subsection{$\mu$-distortion and PTA constraints}
\begin{figure}[t]
  \includegraphics[width=\columnwidth]{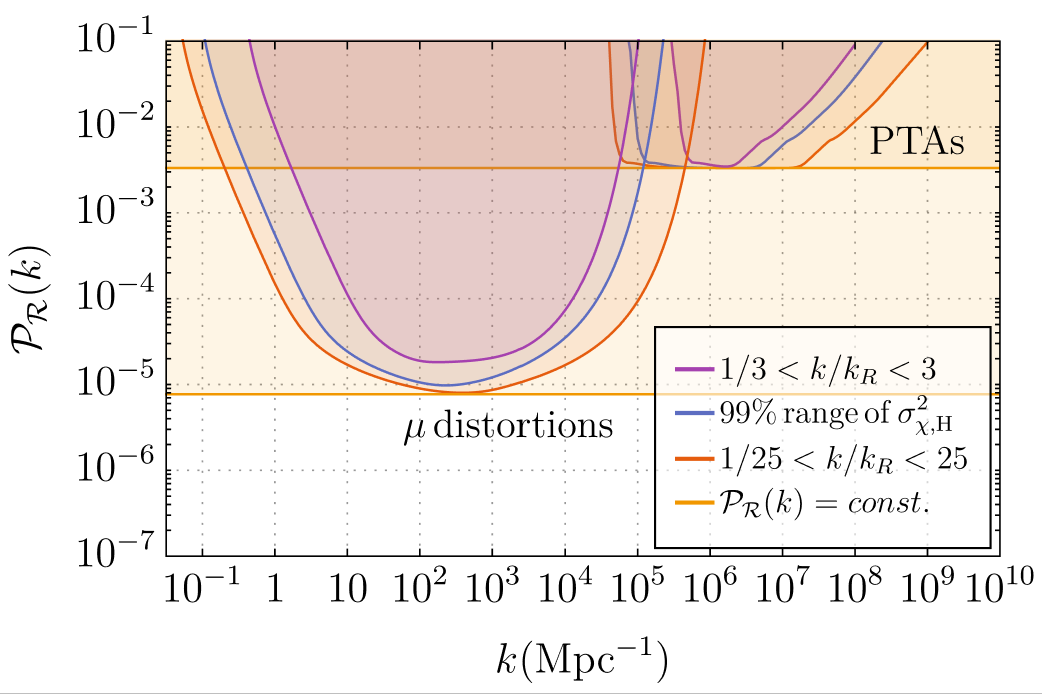}
  \caption{%
  Comparison of the bounds from CMB $\mu$ distortions and PTAs for different integration ranges
  over a scale-invariant spectrum, as indicated. The purple line corresponds to what we adopted in the main
  text, while the blue line corresponds to a range 
  $0.15\lesssim k/k_R\lesssim 12.35$, following Ref.~\cite{Chluba:2012we}.
  }
  \label{fig:range_comparison}
  \end{figure}

In the main text we have pointed out that the CMB spectral distortion and PTA constraints typically 
probe a larger range of scales than UCMHs. In Fig.~\ref{fig:range_comparison} we quantify this statement
by showing the resulting constraints for a spectrum that is scale-invariant, $\pkr=const.$, over various different
ranges of scales $k$ (while assuming $\pkr\simeq0$ outside this range).

We stress that a fully scale-invariant power spectrum, $\pkr=const.$ for {\it all} $k$, 
is largely an academic exercise that we only include here for 
illustration: for normalizations larger than $\sim10^{-9}$, both CMB spectral distortion and 
PTA constraints become irrelevant as 
such a choice is completely excluded by large-scale CMB temperature anisotropy
observations. For direct comparison with the literature~\cite{Chluba:2012we},
we also include a (blue) line corresponding to the range of scales $0.15\lesssim k/k_R\lesssim 12.35$ 
that contribute $99$\% to the mass variance $\sigma_{\chi,H}^2$ given in Eq.~(\ref{eq:sigh}).

In Fig.~\ref{fig:constraints_summary} of the main text, we have presented results for a spectrum 
that is {\it locally} scale-invariant, implicitly assuming that $\pkr$ is non-vanishing only for a relatively small window
in $k$ because the UCMH constraints are anyway only sensitive to scales $k\sim k_R$. Fig.~\ref{fig:range_comparison} 
demonstrates that the UCMH constraints (shown in Fig.~\ref{fig:constraints_summary} in the main text) 
will clearly dominate over both CMB spectral distortion and PTA
constraints even for much `broader' power spectra, with $\pkr\simeq const.$ over a comparatively large range of scales.

\subsection{Constraints from Highly Magnified Stars}
Observations of individual background stars near cluster caustics have opened a novel avenue for constraining compact DM.  
The smooth mass distribution of a galaxy cluster by itself produces a critical curve in the lensed image plane and a 
corresponding fold caustic in the source plane whenever the projected surface density approaches the critical value.  A 
secondary, smaller-scale lens, such as an intracluster (ICL) star or compact object perturbs this macro caustic, creating a web 
of microcaustics around it.  In favourable configurations, where a background star lies close to this combined macro-micro 
caustic, the magnification can reach extreme values, $\mu\sim10^3$ or more, but the 
light curve depends sensitively 
on whether the caustic remains smooth or is fragmented by compact lenses.

The first such constraint was derived from ``Icarus'' (MACS J1149 LS1 at $z=1.49$), whose 
magnification reached 
$\mu \sim 2000$~\cite{Kelly:2017fps,Oguri:2017ock,Diego:2017drh} (with macro-lens magnification $ \mu_t \sim 100$). 
For compact microlenses, $ R_{90} \lesssim \bar{R}_E $, three mass regimes are important: (i) 
for small lenses, the effective Einstein radius ($\bar{R}_E\propto \sqrt{M_{\rm EDO}}$) is smaller than the stellar radius, so the 
star smooths over microcaustics and no constraint arises; (ii) for intermediate masses, the magnification is reduced in a way 
that depends only on the total surface mass density in (all) compact lenses, yielding a constraint on $f_{\rm DM}$ independent of 
$M$; (iii) for large $M$ {(we take $M>2\,{\rm M_\odot}$, following~\cite{Croon:2025yfj}, as this exceeds the mass range expected for intracluster light stars)}, the EDO itself acts as the main lens, and requiring a detectable caustic crossing implies 
$f_{\rm DM} \propto M^{1/3}$, with no constraint for $M \gtrsim 10^4 M_\odot$.

A simplified analytic condition comes from the saturation of the critical curve by microcaustics. The effective Einstein radius of a 
compact lens is $\bar{R}_E \equiv \sqrt{\mu_t} \theta_E D_L$, such that the optical depth is
\begin{equation}
    \Theta = \frac{\Sigma}{M} \pi \bar{R}_E^2,
\end{equation}
where $\Sigma$ is the surface mass density.
A larger $\mu_t$ boosts $\bar{R}_E$, causing the effective lensing cross-section to increase and promoting microcaustic 
overlap even at small $f_{\rm DM}$.
Setting $\Theta=1$ gives an upper limit on the macro-magnification for a given 
$f_{\rm DM} \lesssim f_p = \Sigma/\kappa\Sigma_{\rm crit}$ with 
$\Sigma_{\rm crit} \equiv M/\pi R_{E}^{2} = M/\pi\theta_{E}^{2}D_{L}^{2}$ with the Einstein angle $\theta_{E}$ and $D_{L}$ for 
Icarus given in the main text, and $\kappa \approx 0.83$ the best-fit fraction for the event:\footnote{%
Here, $f_{p}$ quantifies the surface density in compact objects (which includes EDOs as well as ICL stars) normalised by the 
critical surface density. The constraint is conservative when interpreted as a bound on EDOs, since any known stellar 
contribution (e.g., ICL stars) would only enhance the microcaustic density and lead to a lower true value of $f_{\rm DM}$ for 
EDOs.
} 
\begin{equation}
    \mu_{t,\rm max} = \frac{M}{\pi \Sigma R_E^2} \approx 1.2 f_{\rm DM}^{-1}\,,\\
\end{equation}\\
where the last equality takes into account the distances to the lens and source based on the GLAFIC model used for MACS 
J1149 \cite{Oguri:2010rh,Kawamata:2015haa}. 
{For an extended object the effective Einstein radius is reduced relative to the point-like case by an efficiency factor $\varepsilon\le1$, $\bar R_{E,{\rm EDO}} = \varepsilon\,\sqrt{\mu_t}\,\theta_E D_L$, fixed by $\varepsilon^2 = m(\varepsilon\sqrt{\mu_t})$, where $m(\tau)\equiv M(\tau\theta_E)/M$ is the lens-plane-projected mass profile normalised to unity~\cite{Croon:2025yfj}. The saturation condition $\Theta=1$ then generalizes to
\begin{equation}
  \mu_{t,\rm max}\approx 1.2\,f_{\rm DM}^{-1}\,\varepsilon^{-2}\,,
\end{equation}
recovering the point-like result for $\varepsilon\to1$, so that the entire profile dependence of our constraints is carried by $\varepsilon$~\cite{Croon:2025yfj}.}

Crucially, the derivation of the Icarus constraint {also assumes} point sources, an assumption which breaks down when $\bar{R}_E \lesssim R_\star$, as microcaustics are then averaged over by the source's finite size. 
{As a result, no bound applies once the microcaustic size falls below the angular source size, $\theta_E/\sqrt{\mu_t} < \beta_R$ with $\beta_R = R_{\rm source}/D_S$, which sets the lower end of the constrained mass range~\cite{Oguri:2017ock}.}
In part of the regime, complementary constraints were derived based on the expected event 
rate~\cite{Kawai:2024bni}.

Additional events promise to refine or strengthen these constraints. The ``Godzilla'' event in the Sunburst Arc ($z=2.37$) 
reached $\mu \sim 600$ with microcaustic peaks up to a few $10^3$ \cite{Diego:2022mii}; ``Spock'' transients in MACS J0416 
showed $\mu \sim 400$ \cite{rodney2018two}; ``Mothra'' may exceed $\mu \sim 6000$ if its compact perturber is 
confirmed~\cite{Diego:2023qhp}. ``Earendel'' (at $z=6.2$) has been found to have $\mu > 4000$ from JWST 
imaging~\cite{welch2022jwst}.\footnote{%
However, lens models in Ref.~\cite{scofield2025earendel} suggest a lower value of $\mu \approx 43$-$67$. This discrepancy 
highlights the sensitivity of such constraints to lens modeling systematics and underscores the need for continued development 
of cluster mass reconstructions.
}
These increasingly numerous caustic-crossing stars may offer a pathway to probe smaller compact objects than previously 
accessible. Since $\bar{R}_E \propto \sqrt{\mu_t}$, higher-magnification systems enhance sensitivity to lower-mass, as well as 
more extended, lenses.
Future constraints will likely benefit from improved ray-tracing and lens modeling, as well as a statistical 
sample of such events across multiple clusters.

\bibliography{ucmh.bib}

\end{document}